\documentstyle[preprint,revtex]{aps}
\tightenlines
\begin{document}
\draft

\begin{title}
Electronic interactions in fullerene spheres.
\end{title}

\author{F. Guinea \cite{csic}\\}
\begin{instit}
The Harrison M. Randall Laboratory of Physics. \\
The University of Michigan. \\
Ann Arbor, Michigan 48109-1120 USA.
\end{instit}

\author{J. Gonz\'alez \\}
\begin{instit}
Instituto de Estructura de la Materia. \\
Consejo Superior de Investigaciones Cient{\'\i}ficas. \\
Serrano 123, 28006-Madrid. Spain.
\end{instit}

\author{M. A. H. Vozmediano \\}
\begin{instit}
Departamento de F{\'\i}sica Te\'orica. \\
Facultad de Ciencias F{\'\i}sicas.\\
Universidad Complutense.\\
Avda. Complutense s/n 28040 Madrid. Spain.
\end{instit}

\receipt{}

\begin{abstract}

The electron-phonon and Coulomb interactions inC$_{60}$,
and larger fullerene spheres are analyzed.
The coupling between electrons and
intramolecular vibrations give corrections
$\sim 1 - 10$ meV to the electronic
energies for C$_{60}$, and scales as
$R^{-4}$ in larger molecules. The energies
associated with electrostatic interactions
are of order $\sim 1 - 4$ eV, in C$_{60}$
and scale as $R^{-1}$. Charged fullerenes show enhanced
electron-phonon coupling, $\sim 10$ meV,
which scales as $R^{-2}$. Finally, it
is argued that non only C$_{60}^{-}$,
but also C$_{60}^{--}$ are highly polarizable
molecules. The polarizabilities scale as
$R^3$ and $R^4$, respectively. The role
of this large polarizability in mediating
intermolecular interactions is also discussed.

\end{abstract}


\narrowtext

\section{Introduction.}

Fullerene molecules are being discovered in an
atonishing variety of sizes and shapes. The most
studied of them, C$_{60}$ shows a number
of unexpected electronic properties, most
notably the fact that, when combined with many
metallic elements, it gives rise to superconducting
compounds. Recently, a number a other spherical shaped carbon
molecules have been found \cite{fulg1,fulg2}. In the present
work, we discuss the most relevant intaractions which
influence the highest occupied electronic states in these molecules.

The analysis is based on earlier work \cite{GGV1,GGV2},
(hereafter referred as I and II), which
proposed a continuum scheme to analyze the properties of these
molecules. The method makes use of the fact that simple,
long wavelength approximations to the electronic
spectrum of graphite planes  can be made. We showed that
it is possible, in this way, to write down analytical
expressions for the highest occuppied,
and lowest unoccuppied  states of C$_{60}$.
Moreover, the method is more accurate for larger molecules.
In I, a brief outline of its usefulness for the study
of the coupling between electrons and lattice vibrations
was also sketched.

In section II, we describe the most relevant features of the
electronic states of C$_{60}$. The next section deals
with the main properties of the coupling between electrons and phonons
in {\it neutral} molecules. Section IV analyzes the electrostatic
coupling between the electrons. Section V describes the new
electron-phonon couplings which appear in {\it charged}
molecules. In section VI we discuss the electronic polarizabilty
of these molecules, and the interactions which
may be relevant in a crystalline structure. The main
conclusions are analyzed in section VII.

\section{The electronic states of C$_{60}$.}

The electronic properties of conjugated carbon compounds
are determined by the states derived from the unsaturated
$\pi$ orbitals at each carbon atom. The very succesful
H\"uckel model \cite{Huck} uses this fact to describe a
large variety of such materials, from benzene to
conjugated polymers, like polyacetylene, to graphite.
The most important parameter in the model is the
hybridization energy between orbitals placed at
nearest neighbor carbons, which we will take as
$t \sim 2.2$eV. Longer range hybridizations are
neglected.  The remaining three valence electrons
in each C atom give rise to covalent $\sigma$ bonds.
These bonds give rigidity to the lattice. Their spring
constant, $K$, is also well known, which allows
a simple approximation to the vibration spectra
of materials with only carbons, like graphite.
We will assume for the highest frecuency,
$\omega_{opt} = 3 \sqrt{K / M} = 0.20$eV
($M$ is the mass of a C atom). The electronic
states derived from the $\pi$ orbitals are affected
by the lattice vibrations, because they modify the distance
between nearest atoms. This coupling
is described by the variation of the hybridization $t$
with distance. We will take $\partial t / \partial l
= 4.5 \hbox{eV} \hbox{\AA}^{-1}$. There is a greater
controversy about the value of the on site Coulomb repulsion,
when two electrons occupy the same $\pi$ orbital.
In isolated fullerene molecules, metallic screning
is absent, and the long range electrostatic
interactions dominate. We expect also no metallic
screening at distances shorter then the intermolecule
distance in C$_{60}$ crystals. In contrast, screening
effects will be significant in onion like compounds,
where the spheres are stacked inside each other.

The H\"uckel model, when applied to a plane of graphite,
gives rise to two electronic bands which intersect
at two inequivalent points of the Brillouin zone.
Near these points, the spectrum can be written as:
$\epsilon_k = \hbar v_F | \vec{k} |$, where
$v_F = 3 t a / \hbar$, and $a$ is the distance
between nearest carbon atoms. As discussed in I and II,
this dispersion relation can be seen as arising
from an effective long wavelength equation, which
is nothing but the Dirac equation in (2+1) dimensions.
The two inequivalent points of the Brillouin zone
lead to two independent equations. The components
of each spinor correspond to the amplitude of the
wavefunction in each of the two sublattices in
which the honeycomb struvture can be divided.

This long wavelength approximation can be
extended to spherical fullerenes. Naively, one
would expect that the electronic states will
be given by the solutions of the (2+1)
Dirac equation on the surface of a sphere.
The presence of pentagons in the lattice
modifies this picture. When going around one
of them, the two sublattices are interchanged.
One of the components of the spinor associated with
one point of the Brillouin zone is changed into
the other component of the spinor associated with
the second point. This effect can be included by means
of an auxiliary gauge potential, whose
magnetic flux threads each of the pentagons.
Because of the mixing of the two electron
{\it flavors}, this field is non Abelian.
However, as discussed in II, a transformation
is possible, which changes the two
coupled Dirac equations, into two decoupled
ones, which include
an Abelian monopole of opposite charge,
$ Q = \pm 3/2$.

These equations can be solved analytically, by means of
suitably defined angular momentum operators.
The rwsulting spectrum has two degenerate triplets
at $\epsilon = 0$, and a succession of higher
multiplets, with degeneracies $ 2 \times ( 2 j + 1 ), j = 2,3, ...$
and energies $\epsilon_j = \pm v_F / R \sqrt{j(j+1)-2}$.
$R$ is the radius of the molecule.
As extensively analyzed in I and II, this spectrum is
a reasonably good approximation to spherical carbon
fullerenes, becoming more accurate as the number
of atoms in the molecule increases. In particular, for
C$_{60}$, the two degenerate triplets at $\epsilon = 0$
m,entioned above, can be identified with the
lowest low lying molecular orbitals in the molecule.
The fact that these two triplets are split in
C$_{60}$, and that they are not strictly
at $\epsilon = 0$ can be understood as arising from
the next leading corrections to the Dirac
equation. These corrections modify the
spectrum of graphite away from the Fermi points.
They contain additional
derivatives. When included into our effective
Dirac equation for the spherical fullerenes,
they give rise to a coupling between the two triplets,
and an overall shift. The fact that this interpretation is
correct can be seen from the way in which these
terms scale with the radius of the molecule.
The splitting of the two triplets shows the expected
$R^{-2}$ dependence for all sizes, including C$_{60}$.
The behavior of the average energy is more involved,
and the proposed scaling seems to set in only
for large sizes (C$_{540}$ and beyond).

In this way, we complete our description of the two lowest
unoccupied triplets in C$_{60}$ and related molecules.
In lowest order, they
correspond to the states at zero of the Dirac
equation on the surface of a sphere, with an additional
monopole of charge $Q =  \pm 3/2$.
Their wavefunctions are:

\begin{equation}
\begin{array}{rlrl}
\Psi_{+1,a}^{\alpha} &= \sqrt{3 \over{8 \pi}} {\sin}^2 ( {\theta \over 2 } )
e^{i \phi} &\Psi_{+1,a}^{\beta} &= 0 \\
\Psi_{0,a}^{\alpha} &= \sqrt{3 \over{4 \pi}} \sin ( {\theta \over 2} )
\cos ( {\theta \over 2} ) &\Psi_{0,a}^{\beta} &= 0 \\
\Psi_{-1,a}^{\alpha} &= \sqrt{3 \over{8 \pi}} {\cos}^2 ( {\theta \over 2 } )
e^{ - i \phi} &\Psi_{-1,a}^{\beta} &= 0 \\
\Psi_{+1,b}^{\alpha} &= 0 \qquad\qquad
&\Psi_{+1,b}^{\beta} &= \sqrt{3 \over{8 \pi}} {\cos}^2
( {\theta \over 2} ) e^{i \phi} \\
\Psi_{0,b}^{\alpha} &= 0 \qquad\qquad
&\Psi_{0,b}^{\beta} &= - \sqrt{3 \over{4 \pi}}
\sin ( {\theta \over 2} ) \cos ( {\theta \over 2} ) \\
\Psi_{-1,b}^{\alpha} &= 0 \qquad\qquad
&\Psi_{-1,b}^{\beta} &= \sqrt{3 \over{8 \pi}} {\sin}^2
( {\theta \over 2} ) e^{- i \phi} \\
\end{array}
\end{equation}

The superscripts $\alpha$ and $\beta$ refer to the two components
of each spinor, and the subscripts label each of the six
different wavefunctions.

The residual interactions described
earlier can be projected onto the subspace spanned by these six states.
There is a global shift, which scales as
$k_1 t ( R_0 / R )^2$, and an hybridization between
spinors of type $a$ and $b$ with the same angular momentum.
It can be written as $k_2 ( R_0 / R )^2$. $R_0$ is
the radius of C$_{60}$, and the dimensionless
coefficients $k_1$ and $k_2$ can be deduced
from the the spectrum of C$_{60}$. Their values
are 1.53 and 0.09 respectively.

\section{Electron-phonon interaction in neutral molecules.}

We now describe the influence of the various lattice
vibrations in the states given en eq. 1. It is
interesting to note first, that the hopping $t$, and any
modulation of it, couple orbitals in different sublattices.
To lowest order, the orbitals given by (1) represent
electronic states which combine wavefunctions derived from one point
in the Brillouin zone and one sublattice, and wavefunctions
from the other point and the other sublattice.
The hopping does not mix them, and that is why their energies
are at zero. Thus, the only electron-phonon possible,
{\it for these triplets}, arise from the modification
by the phonons, of the next order terms, which were
described before. This effect
was not properly included in I.

The states in equation (1) are delocalized thoroughout the
sphere, and change little from site to site. Hence, they will
only couple to long wavelength phonons. The graphite planes,
which we use as starting point, have two accoustic and two
optical branches near the center of the Brillouin zone,
which we will consider separately.

The accoustical modes are well described by the
elastic theory of spherical shells, as mentioned in I.
They can be parametrized in terms of fields which give
the lattice displacements at each point in the sphere:
$\vec{u} = ( u_r (\theta , \phi ) , u_{\theta} ( \theta ,
\phi ) , u_{\phi} ( \theta , \phi ) )$. On general
grounds, we can clasify them using vector spherical
harmonics for the functions $u_{\theta}$ and $u_{\phi}$,
and ordinary spherical harmonics for $u_r$.
For each value of $l$ and $m$, the calculation of
the eigenmodes is reduced to the diagonalization of a
$3 \times 3$ matrix. The simplest vibration is the breathing
mode, $l = m = 0, u_r = u, u_{\theta} = u_{\phi} = 0$.
This mode simply changes the radius of the sphere.
We know that the lowest order coupling possible is
through the modification of the electronic interactions
within the triplets which decay like $R^{-2}$,
as discussed in the previous section. Thus, its is straightforward
to describe the coupling of the breathing mode
to the lowest lying triplets given in eq. (1).
To leading order in $u / R$, their
mean energy and splitting depend on $u$ as:

\begin{eqnarray}
\bar{\epsilon} &= k_1 {{\partial t} \over { \partial l }} u
{{ {R_0}^2 } \over { R^2 }} \nonumber \\
\Delta \epsilon &= k_2 {{\partial t} \over { \partial l }} u
{{ {R_0}^2 } \over { R^2 }} \nonumber \\
\end{eqnarray}

For a general mode, the coupling to the electrons
can only depend on the strain tensor, given by:

\begin{eqnarray}
u_{\theta,\theta} = &u_r +  \partial_{\theta} u_{\theta}
\nonumber \\
u_{\phi,\phi} = &u_r + {{\partial_{\phi} u_{\phi}}\over
{\sin ( \theta )}} + {{\cos ( \theta )}\over{\sin ( \theta )}}
u_{\theta} \nonumber \\
u_{\theta,\phi} = &\partial_{\theta} u_{\phi} +
{{\partial_{\phi} u_{\theta}}\over{\sin ( \theta )}}
- {{\cos ( \theta )}\over{\sin ( \theta )}} \nonumber \\
\end{eqnarray}

We now make the assumption that the way the strain tensor
changes the electronic levels is through its only scalar
contraction, $u_{\theta , \theta} + u_{\phi , \phi}$.
Physically, this quantity describes the local variation in the area
of the sphere. Generalizing the coupling to the breathing mode,
discussed before, and particularizing in the changes in
$\bar{\epsilon}$, which is the largest effect, we write:

\begin{equation}
{\cal H}_{el-phon (acc)} = 2 k_1 {{\partial t}\over{\partial l}}
\sum_{\alpha , i} \int ( u_{\theta , \theta} + u_{\phi , \phi} )
\Psi_i^{\alpha *} \Psi_i^{\alpha}
\end{equation}

For an elastic vibration with a given $l$ and $m$,
$u_{\theta , \theta} + u_{\phi , \phi} \propto Y_l^m ( \theta , \phi )$.
Given the shape of the wavefunctions (eqn. 1), this result
implies that only modes with $l = 0,1,2$ couple to them.

We can now estimate the change in the electronic levels
due to each lattice mode, using second order perturbation theory.
{}From $ < u^2 > = \hbar / ( 2 M \omega )$ (M is the mass
of the molecule)
the characteristic energy shift induced by a mode of frecuency
$\omega$ is: $( k_1 \partial t / \partial l )^2 / ( 2 M \omega^2 )$.
For C$_{60}$, typical phonon frecuencies are in the range
500 - 1000 cm$^{-1}$. The associated shifts are of order
1 - 5 meV. The dimensionless electron-phonon coupling
parameter, $\lambda$, is $ \le 0.1$ per mode.

In larger molecules, $M \sim R^2$, and $\omega \sim R^{-1}$.
Hence, the shifts will decay as $R^{-4}$, and $\lambda
\sim R^{-3/2}$. Note that the number of accoustic modes which
couple to the states studied here does not scale
with the size of the molecule, because the characteristic
dimension of the electronic wavefunction
is always proportional to the radius of the sphere.

The estimates discussed before point out to a weak coupling
between the accoustical modes and the lowest lying electronic states
of spherical fullerenes.  This effect can be traced back to the
low density of states of graphite near the Fermi level.
For instance, the lowest order coupling between the triplets
in eq (1) and phonons vanishes, but is finite for other
multiplets at higher energies. Furthermore, the extended nature of these
wavefunctions imply that few modes can couple to them.

In molecules with one or two electrons in these states,
a novel effect may arise. As discussed in I, the degeneracy
of the allowed states gives rise to a Berry phase which modifies
the quantization rules for the lattice vibrations. The
combined electron and phonon wavefunction changes its symmetry with
respect to the case of no electron degeneracy. As a consequence,
Raman and infrared modes are exchanged, an effect
which can be verified experimentally.

Finally, we discuss the coupling to optical modes. At each point in the
sphere they can be characterized by its polarization,
longitudinal or transverse, which can be defined
as a two dimensional vector field, $\vec{n} ( \theta , \phi )$.
We will neglect the out of the sphere modes. It is simple
to calculate the coupling between this field and the
spinors which describe the electronic states in
planar graphite. It gives rise to a term in
the hamiltonian:

\begin{equation}
{\cal H}_{el-opt} = \int \bar{\psi} \vec{\sigma} \vec{n} \psi
\end{equation}

where the $\sigma$'s are the Pauli matrices. As in
the case of the accoustical modes, this term does
not modify, to lowest order, the triplets given
in eq. (1).

We can estimate the residual coupling by explicitly
changing the hoppings in the discrete hamiltonian which
describes C$_{60}$, simulating the presence of an
optical phonon. The simplest such mode alters
the pentagon-hexagon and the hexagon-hexagon bonds
in opposite directions. In order for the area of the
sphere to stay constant, the pentagon-hexagon bond
deformation should be close to one half that of the
hexagon-hexagon bonds. The change in the average
energy of the lowest triplets, with respect to
such a modification of the hoppings is
$\partial \bar{\epsilon} / \partial t = 0.9$.
On general grounds we can expect this derivative
to be of order unity. In our continuum formulation,
such coupling can be incorporated into the formalism
by a term:

\begin{equation}
{\cal H}_{el-phon (opt)} = 2 k_3 {{\partial t}\over{\partial l}}
\sum_{\alpha , i} \int \left( \partial_{\theta} n_{\theta} +
{ {\partial_{\phi} n_{\phi}}\over{\sin ( \theta )}}
+ {{\cos ( \theta )}\over{\sin( \theta )}} n_{\theta} \right)
\Psi_i^{\alpha *} \Psi_i^{\alpha}
\end{equation}

where $k_3 = 0.9$.
As in the case of accoustical phonons, this expression
makes clear that only modes with small $l,m$ values couple
to the triplets considered here.

The change in the energy of the electronic states is
$\sim ( k_3 \partial t / \partial l )^2 / ( 2 M
\omega_{opt}^2 )$. Taking $\omega_{opt} = 0.20$eV,
this energy is $\sim 40$meV. The dimensionless
coupling $\lambda$ is $\sim 0.2 - 0.3$ per mode.
These values for the energies and
dimensionless couplings should be independent of
the size of the molecule. Note that $\omega_{opt}$
does not scale with $R$, and $M \sim R^2$.

The combined coupling to all phonons compares well
with more detailed calculations for C$_{60}$\cite{phon1,phon2,phon3}.
Our definition of the
dimensionless coupling $\lambda$ is
$\partial t / \partial l \sqrt{\hbar/(M \omega^2)} / ( \hbar \omega )$,
where $\omega$ is the frecuency of the mode under
consideration. This is equivalent to the formulae
used in \cite{phon1,phon2,phon3}, in the limit when
the coupling is local and the electronic bandwidth tends to zero.

\section{Coulomb interactions in charged fullerenes.}

{}From equation (1) it is simple, although tedious, to calculate
the different Coulomb integrals between orbitals.
The needed integrals are discussed in the appendix.
A variety of estimates are available in the literature,
and our results come close to those reported in ref.
\cite{TDLee}. In fact, if we neglect the splitting
between the two triplets, we obtain,
for C$_{60}^{--}$,  exactly the same succession
of multiplets as in that reference, in what is
called also {\it the continuum approximation}.
We believe that the scheme used in \cite{TDLee}
is equivalent to ours, in that the same wavefunctions are
used. Thus, for C$_{60}^{--}$ the lowest configurations are
two  $^1 s_{+}$ and $^1 s_{-}$ singlets at $3 e^2 / 5 R_0$,
followed by two $^3 p_{+}$  and $^3 p_{-}$ triplets, at
$7 e^2 / 10 R_0$. The splitting between these two
sets of configurations is generated by the second order term
which couples the one electron triplet states,
discussed after eq. 1. An effective hamiltonian
which describes each symmetry ($^1 s$ and $^3 p$)
is given in the appendix.
These terms are also responsible for the splitting
of the triplets in C$_{60}^{-}$.
In the doubly ionized
molecule, however, this term induces a much weaker effect.
The coupling of the two degenerate singlets is through
an intermediate state at $8 e^2 / 5 R_0$, and the coupling between
the triplets is mediated by another triplet at $6 e^2 / 5 R_0$.
In both cases, the splitting goes like $\Delta \epsilon^2  / (e^2 / R_0)$.
For large fullerenes, $\Delta \epsilon \sim R^{-2}$, and this splitting
decays like $R^{-3}$. Further details can be found in the
appendix.

Our method is sufficiently simple to allow us to
calculate electrostatic energies of molecules
with higher charge. For C$_{60}^{---}$, we find, at low energies,
two $^4 s$ multiplets. The electrostatic energy
is $21 e^2 / 10 R_0$. The splitting between these
two configurations is even smaller than in C$_{60}^{--}$,
and goes like $\Delta \epsilon^3 / (e^2 / R_0 )^2 $. For
large fullerenes, this value scales as $R^{-4}$.
In molecules with higher ionization, the splittings
grow again.

Finally, our simple form for the wavefunctions (1), allows us
to characterize the charge distribution of the molecule.
The two orbitals with angular momentum in (1) carry an electric
dipole, $e R_0 / 2$. It is easy to show that the
two $^3 p$ multiplets discussed for C$_{60}^{--}$ carry
the same dipole. The existence of these large dipoles
is an amusing consequence of the fact that the orbitals
given in (1) simulate the state of electrons in a
fictitous magnetic field, described by the monopole.
These orbitals are not chiral invariant. A given
angular momentum also induces an electric dipole.
The terms responsible for the mixing of
these orbitals restore the chiral symmetry.
As  mentioned before, the influence of these terms
is significantly reduced in C$_{60}^{--}$ and
C$_{60}^{---}$, and in larger fullerenes.
The relevance of these dipoles for intermolecular
interactions will be discussed in section VI.

The scale of electrostatic energies, given the radius of
C$_{60}$, $R_0 \sim 3.5 \hbox{\AA}$ is large, $\sim 3$ eV.
We have not investigated to which extent the
level structure of the spheres needs to be modified.
For larger fullerenes, both the separation between
one electron levels and the Coulomb energies
scale in the same way, $\sim R^{-1}$, so both remain
comparable for all sizes.

An alternative method, used to classify the
multiplets of the C$_{60}$ molecule is to
use spherical harmonics and crystal field
terms\cite{Auer,Dress}. This scheme is equivalent
to use a continuum approximation for the
bottom of the graphite band (the point $\vec{k} = 0$
in the Brillouin zone), and to fill all the levels
up to the required charge state. Our method
has the right multiplicity and symmetry for the
relevant orbitals near the neutral state,
as it starts from an approximation which
describes graphite near its Fermi energy.
Moreover, the separation between levels,
which in other schemes arises from
crystal field effects, appears in a natural
way, and its scaling as function of size
can be easily investigated. In that respect,
we differ somewhat from the assumptions
about scaling in ref.\cite{Dress}.

In onion shaped materials, the energy levels
are affected by the tunneling between contiguous
layers. The energy of these processes is,
in graphite, $\sim 0.3$ eV \cite{grap}. The Coulomb
interactions will be more drastically changed,
by the appearance of screening effects.

\section{Electron-phonon interaction in
charged fullerenes.}

In charged fullerenes, the deformations of the lattice
induce rearrangements in the charge distribution,
which modify the Coulomb energies. As discussed in the
previous section, these energies are important,
so that this effect needs to be taken into account.
This effect is unique to systems like the doped fullerenes,
which combine localized electronic states and
inhomogeneous charge distributions. Hence,
other conjugated compounds,
like graphite intercalates, offer no clue to
the strength of this coupling.

These modifications in the charges
over scales comparable to the size of the molecule
can only be induced by long wavelength
accoustical phonons. Optical modes redistribute
the charge within each unit cell, but do not alter
the overall charge distribution.

The leading effect of a deformation of the lattice is to
change locally the area element. The charge density
in that region is then modified. Again, we can discuss
this effect taking the simplest lattice vibration, the breathing
mode as an example. The area of the sphere is expended or
contracted, and the Coulomb energies change due to the
variation in the radius of the sphere. There is no effect
in C$_{60}^{-}$. In C$_{60}^{--}$, the change of energy associated
to a vibration of amplitude $u$ is $\sim e^2 u / R_0^2$.
For a general mode, the charge density at a given place
decays as the area element increases. The local expansion
induced by a given mode is $u_{\theta , \theta} +
u_{\phi , \phi}$. Thus, we have to insert
this expression in the Coulomb integrals
between different orbitals. As in the case of the
standard coupling to accoustical phonons, this
result imply that only vibrations with
$l \le 4$ can couple to the triplets given by eq. 1.

We can make an estimate of the energy shift in the electronic
levels in the same way as in the case of the
modification in the hybridization due to the vibration.
The only difference is that the factor $( k_1 \partial t / \partial l )^2$
has to be replaced by $( e^2 / R_0^2 )^2$.  The second
factor is larger by a factor of 8 - 10. Thus, the energy
scales at wich this coupling can play a role are $\sim
20 - 40$meV. The values of the dimensionless
constant $\lambda$ will become $\sim 0.8 - 1$.
These interactions are significantly larger than
the standard electron-phonon couplings. The energies
should scale as $R^{-4}$ for large molecules, and the
values of $\lambda$ as $R^{-3/2}$.

\section{Interactions between different fullerene
molecules.}

So far, we have discussed the various electronic interactions
within a given fullerene molecule. It is interesting
to analyze which ones may play a role when
different molecules are close to each other.
As the most important influence that we have found
is due to the Coulomb interaction, we
will analyze further its role in this case.

We will assume that charge that the fullerenes may
have is compensated by some neutralizing background,
so that the only residual interaction is due to the Van der Waals
forces.  Then, the relevant magnitude is the
electrical polarizability of the molecules.
As mentioned in section IV, C$_{60}^{-}$ is highly polarizable,
a result also discussed in ref.\cite{TDLee}.
Also, C$_{60}^{--}$ has a large polarizability, if the molecule is
in a $^3 p$ state. An estimate of the polarizabilities
of various charged states of fullerenes is given in table I.
These polarizabilities are due to the extra electrons
which are present in charged C$_{60}$ molecules. The polarization
of the closed shell in neutral C$_{60}$ has been estimated
\cite{pol} to be $\sim 80 \hbox{\AA}^3 \sim 300$ a. u., that is,
significantly smaller than the values we find
for C$_{60}^-$ and C$_{60}^{--}$.

The high polarizability of C$_{60}^{--}$ can play a role,
if the interaction between neighboring molecules overcomes
the energy difference betwwen the $^1 s$ and the $^3 p$
configurations. This difference is $e^2 / ( 10 R_0 )
\sim 0.5$eV, as discussed in section IV. The van der Waals
interaction between two molecules at distance $D$ is:
$\sim e^4 R_0^4 / ( 4 D^6 \epsilon_p )$, where
$\epsilon_p$ is the splitting between the two $^3 p$
configurations, $\sim 2 \Delta \epsilon^2 / ( e^2 / R_0 )$
(see the appendix for details of the calculation).
Hence, the van der Waals energy is
$\sim  5 e^6 R_0^3 / ( 4 D^6 \Delta \epsilon^2 )$,
where $\Delta \epsilon \sim 0.40$eV, as discussed in section 2.
Even for $D = 4 R_0$, this energy is $\sim 0.2$eV,
and increases rapidly at shorter distances.
Thus, it is not unconceivable that C$_{60}^{--}$ will exist in its
$^3 p$ state in a crystal, and neighboring molecules
will interact strongly through their mutual dipoles.

\section{Conclusions.}

We have investigated the various interactions which may
influence the electronic properties of a fullerene molecule.
The relevant energies, and the scaling as function of molecule
size are given in table II.

Our results suggest that the most relevant effects will arise
from the Coulomb interactions. The weakness of the electron phonon
coupling can be related to a similar effect in graphite,
due to the vanishing density of states near the Fermi {\it points}.
Phonons can only play a significant role in charged
molecules, where they modify the charge distribution.

On the other hand, long range Coulomb interactions have no
counterpart in other conjugated compounds. The most
studied charged (through doping) carbon systems,
graphite intercalation compounds and poliacetylene,
do not show these features. In both cases, the
relevant electronic states are delocalized, and are
not susceptible to charging effects.

We have pursued further the role of Coulomb interactions
in crystalline systems. The most striking
effect that we have found is the extremely
high polarizability of C$_{60}^{--}$ molecules.
This property leads to significant
intermolecular interactions at distances
$\sim 3 - 4$ times the radius of the molecule.
The scale of these couplings $\sim 0.3 - 3$eV,
can easily overcome most other efffects in
a doped C$_{60}$ crystal, and may be even responsible for the
unusual dipolar moments found in neutral
C$_{60}$\cite{dipole}. Its role in other
properties, like superconductivity, is being
further investigated.

\unletteredappendix{Calculation of Coulomb integrals.}

The Coulomb integrals required to evaluate the energies
of the various configurations of charged C$_{60}$ can
be obtained from the wavefunctions given in eq. (1).
We denote the six orbitals in that equation as:
$ | a ,+1 > , | a , 0 > , | a , -1 > , | b , +1 > , | b , 0 > $
and $ | b , -1 >$, where we are omitting spin indexes.
 The Coulomb potential can be
expanded in spherical harmonics,
$1 / | \vec{r}_1 - \vec{r}_2 | = 4  \pi \sum_{l} 1 / ( 2 l + 1 )
\sum_m Y_l^{-m} ( \hat{r}_1 ) Y_l^m ( \hat{r}_2 )$,
where we are using the fact that
$ | \vec{r}_1 | = | \vec{r}_2 | = 1$, and we use units such that
$R_0 = 1$. Then, the seven different Coulomb integrals are
(in units of $e^2 / R_0$):

\begin{equation}
\begin{array}{ccc}
v_{a+1,a+1;a+1,a+1} = v_{a+1,a+1;b-1,b-1} &= &{{63}\over{50}} \\
v_{a+1,a+1;a0,a0} = v_{a+1,a+1;b0,b0} &= &{{49}\over{50}} \\
v_{a+1,a+1;a-1,a-1} = v_{a+1,a+1;b+1,b+1} &= &{{19}\over{25}} \\
v_{a+1,a0;a+1,a0} &= &{{7}\over{25}} \\
v_{a+1,a0;b+1,b0} &= &- {{11}\over{50}} \\
v_{a+1,a-1;a+1,a-1} &= &{{3}\over{50}} \\
v_{a0,a0;a0,a0} = v_{a0,a0;b0,b0} &= &{{26}\over{25}} \nonumber \\
\end{array}
\end{equation}

The remaining integrals can be obtained by exchanging the $a$ and $b$
indices.

As mentioned in section IV, this procedure leads to the
multiplet energies for C$_{60}^{--}$ discussed in\cite{TDLee}.
The wavefunctions with $^1 s$ symmetry are:

\begin{eqnarray}
| 1 > &= {1\over{\sqrt{3}}} \big( | a +1 \uparrow , a -1 \downarrow >
- | a 0 \uparrow , a 0 \downarrow > + | a -1 \uparrow ,
a +1 \downarrow > \big) \nonumber \\
| 2 > &= {1 \over {\sqrt{6}}} \big( | a +1 \uparrow , b -1 \downarrow >
- | a  +1 \downarrow , b -1 \uparrow >
+ | a -1 \uparrow , b +1 \downarrow > \nonumber \\
&- | a -1 \downarrow , b +1 \uparrow > - | a 0 \uparrow , b 0 \downarrow >
+ | a 0 \downarrow , b 0 \uparrow > \nonumber \\
| 3 > &= {1\over{\sqrt{3}}} \big( | b +1 \uparrow , b -1 \downarrow >
- | b 0 \uparrow , b 0 \downarrow > + | b -1 \uparrow ,
b +1 \downarrow > \big) \nonumber \\
\end{eqnarray}

These three states are mixed by the term in the C$_{60}$
hamiltonian which gives rise to the splitting of the
two triplets, $\Delta \epsilon$. Thus, we end up with
the matrix:

\begin{equation}
{\cal H}_{^1 s} =
\left( \begin{array}{ccc}
{{3 e^2}\over{5 R_0}} &{{\Delta \epsilon}\over{\sqrt{2}}} &0
\\{{\Delta \epsilon}\over{\sqrt{2}}} &{{8 e^2}\over{ 5 R_0}}
&{{\Delta \epsilon}\over{\sqrt{2}}} \\
0 &{{\Delta \epsilon}\over{\sqrt{2}}} &{{3 e^2}\over{5 R_0}}
\end{array} \right)
\end{equation}

Similarly, the states with $^3 p$ symmetry are:

\begin{eqnarray}
| 1 > &= | a +1 \uparrow , a 0 \uparrow > \nonumber \\
| 2 > &= {1 \over{\sqrt{2}}} \big( | a +1 \uparrow , b 0 \uparrow >
+ | b +1 \uparrow , a 0 \uparrow > \big) \nonumber \\
| 3 > &= | b +1 \uparrow , b 0 \uparrow > \nonumber \\
\end{eqnarray}

and the corresponding effective hamiltonian is:

\begin{equation}
{\cal H}_{^3 p} =
\left( \begin{array}{ccc}
{{7 e^2}\over{10 R_0}} &{{\Delta \epsilon}\over{\sqrt{2}}} &0 \\
{{\Delta \epsilon}\over{\sqrt{2}}} &{{6 e^2}\over{ 5 R_0}}
&{{\Delta \epsilon}\over{\sqrt{2}}} \\
0 &{{\Delta \epsilon}\over{\sqrt{2}}} &{{7 e^2}\over{10 R_0}}
\end{array} \right)
\end{equation}

For the purposes of the discussion in section VI, it is important
to realize that the states $| 1 >$ and $| 3 >$ in (10) carry an
electric dipole, $\pm e R_0 / 2$.

\newpage
\begin{table}

\caption{Typical energy scales associated with the various interactions
discussed in the text, for C$_{60}$, and scaling of each of them as
function of molecule radius, for larger fullerenes.}

\begin{tabular}{c|c|c}

    Interaction    & Energy & Scaling \\
\tableline
    Electron-phonon (accoustic)             & 0.004 eV & $R^{-4}$   \\
    Electron-phonon (optical)               & 0.04 eV     & $R^0$    \\
    Coulomb                                 & 3 - 6 eV   & $R^{-1}$  \\
    Electron-phonon (in charged molecules)  & 0.1 eV     & $R^{-4}$  \\
\end{tabular}
\end{table}

\begin{table}
\caption{Polarizability (of
the valence electrons) in different charge states of C$_{60}$
(in atomic units), and scaling behavior for larger molecules.}

\begin{tabular}{c|c|c}

Charge state & Polarizability (in a. u.) & Scaling \\
\tableline
Neutral	( $^1 s$ )	& 0		& $R^0$ \\
1 e$^-$	( $^2 p$ )	& 10$^3$	& $R^4$ \\
2 e$^-$ ( $^1 s$ )	& 0		& $R^0$ \\
2 e$^-$ ( $^3 p$ )	& 10$^5$	& $R^5$ \\
3 e$^-$ ( $^4 s$ )	& 0		& $R^0$ \\
\end{tabular}
\end{table}
\end{document}